\documentstyle[12pt]{article}
\begin{document}
\title{Quantum field theory can be UV finite if it is taken as 
effective one}
\author{Jifeng Yang$^*$ \\
Department of Physics and School of Management, \\
Fudan University, Shanghai, 200433, P. R. China}
\footnotetext{$^*$ E-mail:jfyang@fudan.edu.cn}
\date{\today}
\maketitle
\begin{abstract}
This is a further explanation of a recent approach proposed by the
author (hep-th/9708104, Ref. \cite {YYY}, that is somewhat sketchy) 
for any ordinary QFT (whether renormalizable or not) in any
space-time dimension. We discussed the physical motivations of the 
new approach and its efficiency when compared to the existent 
renormalization approaches. Some other important issues related
are briefly touched.
\end{abstract}

PACS number(s): 11.10.-z; 11.10.Gh; 11.15.-q; 11.15. Bt
\vspace{2.5cm}
\newpage
It is known to all that the old frameworks of renormalization 
first invoke UV infinities and then try to find some doubtful 
'operation' to remove them in order to predict the obviously finite
world \cite {Jac}. The worse is, one has to find a regularization 
(Reg) first in the intermediate stage of the framework without 
appreciating the physical implication of this technical necessity.
In short, the difficulty is inevitable if one hold the present 
formulation of QFTs to be complete and elementary. The necessity 
of introducing a regularization (in whatever way \cite {DR}) 
itself means already that the present formulation of QFTs is not 
a complete or fundamental one.

Now, that a fundamental theory underlies all the ordinary QFTs 
with the latter ones being various low energy effective theories 
has become a standard point of view \cite {wein}. But as far as 
the author knows, we are still lacking a formulation that can 
yield finite results in a natural way (without invoking \it ad 
hoc \rm regularizations and divergences) that fully makes use of 
the standard point of view. A new approach is proposed in 
Ref. \cite {YYY} that fully exhibits the power of the standard 
point of view if one uses it appropriately. (The Wilsonian 
approach \cite {wil} which works perfectly in the context of 
critical phenomena, is questionable if one applies it to all 
ordinary QFTs in the original sense as then it can only deal 
with the renormalizable ones in an \it ad hoc \rm way (see, \cite 
{pol}, Eq.(18)). While, our new approach is rather simple and 
applies to all QFT models and all space-time dimensions.)

Let me repeat some part of Ref. \cite {YYY} and add some 
discussions and explanations where necessary.

First the standard point of view is restated as follows: suppose, 
the true complete theory underlying the present QFTs is found, 
\it (1) it must be well defined in every aspects and always yield 
physically sound (finite, of course) predictions in any energy 
range, at least for those ranges supposed to be well described by 
present QFTs; (2) it must have been characterized by certain new 
parameters dominant in the extremely high energy end ( in order 
to define the theory completely and unambiguously); (3) all the 
objects described by the Feynman Amplitudes (FAs) or other 
quantities (perturbative or nonperturbative ones) from the 
present formulation of QFTs should first be derived or calculated 
from the underlying theory with certain limit operation about its 
fundamental parameters afterwards as we are presently in a "low 
energy" phase. \rm Then we can identify the origin of the UV 
infinities: ill-defined (or divergent) FAs (or other objects) 
directly obtained from the present formulation of QFTs are 
consequences of illegal operations on the corresponding 
"amplitudes" from the underlying theory. 

In formula, if the integrand  
$ f(\{Q_{i}\}, \{p_{j}\},\{m_{k}\})$ of an 
ill-defined FA corresponds to the integrand 
$\bar{f} (\{Q_{i}\},\{p_{j}\},\{m_{k}\}; \{{\sigma}_{l}\})$ 
from the underlying theory with 
$\{Q_{i}\}, \{p_{j}\},\{m_{k}\},\{{\sigma}_{l}\}$ being 
respectively loop momenta, external momenta, masses and the 
fundamental parameters in the underlying theory, then
\begin{eqnarray}
&\Gamma^{0}& (\{p_{j}\},\{m_{k}\}) = {\bf L}_{\{\sigma\}}
\bar{\Gamma} (\{p_{j}\},\{m_{k}\};\{\sigma_{l}\})\nonumber \\
&= &{\bf L}_{\{\sigma\}} \int \Pi_{i} d^{n}Q_{i} \bar{f}
(\{Q_{i}\},\{p_{j}\},\{m_{k}\};\{\sigma_{l}\})\nonumber \\
&\neq& \int \Pi_{i}d^{n} Q_{i} {\bf L}_{\{\sigma\}} \bar{f}
 (\{Q_{i}\},\{p_{j}\},\{m_{k}\};\{\sigma_{l}\})\nonumber \\
&=& \int \Pi_{i} d^{n}Q_{i} f(\{Q_{i}\},\{p_{j}\},\{m_{k}\}),
\end{eqnarray}
where $\Gamma^{0}$ and $\bar{\Gamma}$ are well-defined 
(finite), the symbol ${\bf L}_{\{\sigma\}}$ denotes the limits 
operations and $n$ denotes space-time dimension. That means, 
${\bf L}_{\{\sigma\}}$ and $\int \Pi_{i} d^{n}Q_{i}$ do not
commute on all the integrands $\bar{f}(...)$, i.e., the 
commutator 
\begin{equation}
\delta_{\{\sigma\}}= \left [ {\bf L}_{\{\sigma\}},
\int \Pi_{i} d^{n}Q_{i} \right ]
\end{equation}
only vanish identically for convergent or well-defined FAs, 
otherwise we meet troubles: divergence or ill-definedness in FAs. 
Or, for an ill-defined objects from ordinary QFT
$\Sigma_{\{j\}} \{ F(...,\{m\},\{\alpha\}, \{j\} ) \}$ where
$\Sigma_{\{j\}}$ refers to the general operation of summing
over the intermediate states or the virtual processes, its 
correct formulation from the underlying theory should be 
${\bf L}_{\{\sigma\}} (\Sigma_{\{j\}} ( \bar{F} (\cdot, \{j\}; \{\sigma\})))$
which should be  well-defined now. 
Then the ill-definedness of the former tells
that the exchange of the two operations ${\bf L}_{\{\sigma\}}$
and $\Sigma_{\{j\}}$ is illegal,
\begin{equation}
\left [ {\bf L}_{\{\sigma\}}, \Sigma_{\{j\}} \right ]
\neq 0.
\end{equation}
Note that the limit operation operated after the internal 
integration(s) may yield some local terms with finite and
definite constants that reflecting the influence of the short
distance theory or structures on the low energy physics. This will
be picked up later where its importance is addressed.

Now, as a by-product, we can see that a Reg amounts to a 
necessary but  "artificial substitute" for the inaccessible 
"truth", the highest energy structures of the world, which \it
may \rm still be burdened by divergences apart from the side 
effects like violations of symmetries of the original theory as 
cost \cite {DR}.

In principle, everything of the effective QFTs should be well-
defined subsets in the underlying theory, they are the correct, 
finite and at the same time unambiguous expressions of all the
objects that will appear in low energy ranges. Or we can
calculate a subset of functionals from the underlying theory 
that will finally give us the well-defined 1PI Green functions'
generating functional or well-defined path integrals (surely
different from their present forms which is ill-defined) for the 
effective theories derived from the underlying theory. We can 
of course obtain the action functionals (or the Lagrangians) for 
the ordinary QFTs (now as effective ones) up to equivalence. 
\it But all these are correctly obtained only if we apply the 
limit operation after all other operations (especially the 
internal integrations or the summations over the intermediate 
states) have been done, i.e., only if we have followed the 
correct order. If one first obtain the action for an effective
theory before any internal integration is done by applying the 
limit operation first, then one goes back to the present 
formulation of QFTs, and ill-definedness shows up. \rm Thus, it
is not correct to calculate quantum corrections directly from
the effective actions (via present formulation of quantizations).
In other cases, one can not claim that an 'underlying theory' is 
final and well-defined merely because it can yield finite low
energy actions for the phenomenologically established QFTs.
One should check whether other quantities obtained from his
claimed-to-be-final theory is well-defined when the low energy
limit is taken.

But the underlying theory or the expressions 
$\bar{f}(...;\{\sigma_{l}\})$ are unknown by now, we have 
to find a natural way to approach the truth (without introducing 
any \it ad hoc \rm or artificial 'deformations')
$\Gamma^{0}(\{p_{j}\},\{m_{k}\})$'s. In the following, we will 
demonstrate a new and tractable way to achieve this goal which 
is different from any existent methods.

We will start from the following fact (making use of the well 
known fact that differentiation wrt mass or external momenta
can reduce the divergence degree \cite {CK}) for 1-loop case 
ill-defined FAs to try to find finite expressions,
\begin{equation}
\int d^{n}Q \left ({\partial}_{p_{j}} \right )^{\omega} 
f(Q,\{p_{j}\},\{m_{k}\})= 
\left ( {\partial}_{p_{j}} \right )^{\omega} \Gamma^{0} 
(\{p_{j}\},\{m_{k}\}),
\end{equation}
with $\omega-1$ being the usual superficial divergence degree of
$\int d^{n}Q f (Q,\{p_{j}\},\{m_{k}\})$ so that the lhs of Eq(4) 
exists (finite) and $\left ({\partial}_{p_{j}} \right )^{\omega}$ 
denoting differentiation's wrt $\{p_{j}\}$'s.  For the simple 
proof of this fact please see Ref. \cite {YYY}.

The rhs of Eq(4) can be found as the lhs now exists as a
nonpolynomial (nonlocal) function of external momenta and masses.
To find $\Gamma^{0} (\{p_{j}\},\{m_{k}\})$, we integrate both 
sides of Eq(4) wrt the external momenta "$\omega$" times 
indefinitely and arrive at the following expressions
\begin{eqnarray}
& &\left (\int_{{p}}\right )^{\omega}
 \left [ ({\partial}_{{p}})^{\omega} \Gamma^{0} 
(\{p_{j}\},\{m_{k}\}) \right ] = \Gamma^{0} (\{p_{j}\},\{m_{k}\}) 
 +  N^{\omega} (\{p_{j}\},\{c_{\omega}\}) \nonumber \\
&=& \Gamma_{npl} (\{p_{j}\},\{m_{k}\}) + N^{\omega} 
(\{p_{j}\}, \{C_{\omega}\}) 
\end{eqnarray}
with $\{c_{\omega}\}$ and $\{C_{\omega}\}$ being arbitrary 
constant coefficients of an $\omega-1$ order polynomial in 
external momenta $N^{\omega}$ and 
$\Gamma_{npl} (\{p_{j}\},\{m_{k}\})$ being a definite 
nonpolynomial function of momenta and masses \cite {YP}. 
Evidently $\Gamma^{0} (\{p_{j}\},\{m_{k}\}) $ is not uniquely 
determined (within conventional QFTs) at this stage. That the 
true expression
\begin{equation}
 \Gamma^{0} (\{p_{j}\},\{m_{k}\}) = \Gamma_{npl} 
(\{p_{j}\},\{m_{k}\}) + N^{\omega} 
(\{p_{j}\},\{\bar{c}_{\omega}\}) , \ \ \ \bar{c}_{\omega}= 
C_{\omega}-c_{\omega}
\end{equation}
contains a definite polynomial part (unknown yet) implies that 
it should have come from the low energy limit operation on 
$\bar{\Gamma} (\{p_{j}\},\{m_{k}\};\{\sigma_{l}\})$ (see Eq(1)) 
as the usual convolution integration can not yield a polynomial 
part, an indication of the incompleteness (or ill-definedness)
of the present QFTs.

We can also take the above procedure as a natural way of 
rectifying the ill-defined FAs that "replaces" them with the 
expressions like the rhs of Eq.(5), i.e., 
\begin{equation}
\int d^{n}Q f (Q,\{p_{j}\},\{m_{k}\}) >=< \Gamma_{npl} 
(\{p_{j}\},\{m_{k}\}) + N^{\omega} (\{p_{j}\}, \{C_{\omega}\})
\end{equation}
with "$>=<$" indicating that rhs represents lhs \cite {YYY,YP}. 
That the ambiguities reside only in the local part means the QFTs 
are indeed effective low energy ones.

To find the $ {\bar{c}_{\omega}}$'s in Eq.(6) we need inputs from 
the physical properties of the system ( such as symmetries, 
invariances, unitarity of scattering matrix and reasonable 
behavior of differential cross-sections) and a complete set of 
data from experiments \cite {CK,LL} (if we can derive them from 
the underlying theory all these requirements would be 
automatically fulfilled) as \sl physics determine everything 
after all. \rm  In other words, all the ambiguities should be
'fixed' in this way. Note that this is a principle independent of 
interaction models and space-time dimensions, i.e., we can 
calculate the quantum corrections in any model (whatever its 
'renormalizabilty' is) provided the definitions can be 
consistently and effectively done. Similar approach had been 
adopted by Llewellyn Smith to fix ambiguities on Lagrangian level 
by imposing high energy symmetry, etc. on relevant quantities 
\cite {LL}. For the use of later discussion, I would like to 
elaborate on the implications of the constants. As we have seen, 
the $ \bar{c}_{\omega} $ 's arise in fact from the low energy 
limit operation on the objects already calculated in the 
underlying theory, they are uniquely defined given any set of 
specific low energy parameters (often as Lagrangian or 
Hamiltonian parameters) up to possible reparametrization 
invariance. The choosing of renormalization conditions in the
old renormalization procedure just corresponds to this important 
step in our present formulation for the 'renormalizable' models. 
It is easy to see that if one defines the $\bar{c}_{\omega} $'s 
differently (chooses the ren. conditions differently in the old 
ren. theory) modular the reparametrization equivalence, then the 
physical contents of the corresponding (effective) theory hence 
defined would necessarily be different, or even could not 
describe relevant low energy physics. On the other hand, if one 
think of different definitions as the limits of different 
underlying theories, then it is clear that the low energy 
effective theories can not be independent of the underlying 
theory(s), i.e., \it the underlying theory(s) stipulates or 
defines the effective ones through these constants though the 
fundamental parameters characterizing the underlying theory 
do not explicitly appear in the latter ones. \rm Thus, our 
approach naturally highlights the step of defining these 
constants, while all the usual approaches seemed to have 
failed to appreciate this important aspect. 

The generalization of the treatment of the 1-loop case to the 
multi-loop case is straightforward and simple in concept, we will
report it in another paper forthcoming where many conventional
subtleties in loop momenta integrations will be elucidated
in our new approach \cite {Y1}. It is time now to present a 
critical observation on the multi-loop 1PI FAs containing 
ill-definedness (in the following discussion we should always bear 
in mind that for any FA there is a unique well defined "original" 
counterpart in the underlying theory): \it different treatment 
(e.g., various parametrization operations on such FAs ) would 
produce different results (carrying different form of ambiguities 
or divergence's). \rm (It is a serious challenge for the 
conventional renormalization as choosing the treatments 
arbitrarily would make it impossible to define the counterterms 
consistently at all.) This is ridiculous as these operations (not 
affecting the structures of the amplitudes at all) should be of 
no concern at all. With our preparations above we can easily find 
the origin of this trouble as identified above: QFT "has 
unconsciously performed some illegal (or unjustified) operations 
first". Then the solution follows immediately where a new 
mechanism is used.

For convenience we divide all the graphs (or FAs ) into three 
classes: (A) overall divergent ones; (B) overall convergent ones 
containing ill-defined subgraphs; and (C) the rest, totally well 
defined graphs. We need to resolve all kind of ambiguities in 
classes (A) and (B). Note that any subgraph ill-definedness can 
be treated similarly as in Eq(7) including the overlapping 
divergent graphs \cite {Y1}). First let us look at class (B). 
For a graph in this class, one would encounter nonlocal 
ambiguities due to the subgragh ill-definedness. While such 
graphs must correspond to certain physical processes as they 
carry more external lines, thus, the ambiguities in their 
nonlocal expressions will in principle be fixed or removed by 
relevant experimental data, that is, \it the ambiguities in the 
subgraphs are also constrained by "other graphs". \rm So, with the 
experimental data, the nonlocal ambiguities (from the local
ambiguities of the subgraphs in fact) are in principle completely
fixed or removed. 

To solve the problem with class (A), we note that class (A) can 
all be mapped into class (B) as subgraphs of the latter, then the 
resolution of the ambiguities in class (A) follows immediately. Thus, 
to our surprise, in this simple approach incorporating the Feynman 
graph structures, all the potential ambiguities or divergence's 
should not materialize at all. (\it This fact, in our eyes, 
underlies the magnificent success of QED traditionally treated with 
some mysterious procedures. \rm Now the unreasonable procedures can 
be replaced by our approach to be standardized later.) The important 
thing is this resolution is valid for the complete theory, that is, 
a nonperturbative property rather than a perturbative one. 

Here is a new question: as the ambiguities in one subgraph can in 
principle be fixed or removed through restrictions from 
different overall convergent graphs or from different experimental 
inputs, then, can these "definitions" be consistently done? The 
answer will certainly depends on model structures, then a new 
classification for the QFT models for certain energy ranges based on 
such consistency shows up : category one ( $FT_{I}$ here after) with 
consistent "definitions" implementable, category two ($FT_{II}$) 
without such consistency. Of course $FT_{I}$ interests us most, but 
as the energy range of concern extends upward, the set $FT_{I}$ will 
"shrink" while the set $FT_{II}$ will swell. The final outcome of 
this "move", if accessible at all, should be the final underlying 
theory unique up to equivalence (like the present situation in 
superstring theories \cite {JH} somehow), being or not being a
field theory \cite {Jac}. As for the relation between this 
classification and that judged by renormalizability, we can claim
rigorously before further investigations is done. Intuitively QED, 
etc. seem to belong to $FT_{I}$.

Now let us discuss a formulation based on Wilson's picture 
\cite {pol}. We note that Wilson's picture is basically the same 
as the one we used as standard point of view (term as a natural 
postulate in \cite {YYY}). But it is crucial to note that the 
formulation of Ref. \cite {pol} is based on such an interpretation 
of the Wilsonian picture, i.e., the content of the low energy 
physics is independent of the short distance theory (or the 
underlying theory) up to parameter redefinition effects. 
However, from our discussions above, this is an ad hoc assumption 
as the renormalization conditions affect physics and the 
independence of the low energy theories upon the short-distance 
theory scales (correspond somehow to the $\{\sigma\}$ in our
formulation) does not necessarily mean that the effective theories 
are independent of the renormalization conditions. The only 
possibility that it may work is that one considers a rather 
special set of theories, i.e., the conventionally 'renormalizable'
ones given that one has correctly chosen the renormalization 
conditions. This automatically leads to the method's incapability of 
dealing with the conventionally so-called unrenormalizable 
theories (that are in fact physically interested like gravity) 
while the current trend cares a lot about the 'unrenormalizable' 
ones \cite {gom}. The renormalization group (RG) invariance followed 
from this ad hoc interpretation, if not effected as the 
reparametrization invariance of the low energy physics system,
is in question as the real scale transformation property of 
the system should not be effected in this way. The worse is, the 
reparametrization invariance is not generally guaranteed for the 
whole theoretical contents of certain kind of models (they are 
only implementable for the 1-P-I Green functions for the 
renormalizable theories), let alone for the other kind of 
('unrenormalizable') models. We would like to point out here that 
though the usual arguments for the renormalization group equations 
break down, the renormalization-group-like equations can still be 
derived in certain cases as the real property of some physical 
systems \cite {JY} and it is related to the IR properties of the 
effective theories and the original application of Wilson's RG in
critical phenomena \cite{wil}.

For the infrared (IR) problem, we do not elaborate any more as it 
is given in \cite {YYY}. We only point out here that the IR problem 
for gauge theories is in fact due to the degeneracy of charge 
particle states "wearing" soft boson clouds \cite {Kino}
and its deeper origin is shown to be the conflict between gauge 
symmetry and Lorentz invariance \cite {Haag}. Hence the IR issue 
would contribute something nontrivial to the physical constraints
on the set $FT_{I}$. Besides this, our recent works showed that 
a kind of an unambiguous IR singular term like 
$p_{\alpha}...q_{\beta}...\displaystyle \frac{k_{\mu}k_{\nu}}{k^2} 
(k=p+q)$ originates anomalies ( chiral and trace) no matter how 
one defines the ambiguous polynomial (or what Reg's are employed 
\cite {YP,Acta}), i.e., anomalies arise from unambiguous IR 
(physical) structures rather than from regularization effects or the
inevitability of anomaly (chiral or trace) for the present matter and 
interaction contents is inherent in the low energy theories and 
independent of the underlying theory. 

We want to point out another observation from our approach that
the conventional quantization procedure of fields is now subject to 
question. Especially, the elementary commutator for a field 
(fermionic or bosonic) and it conjugate, if calculated
(or formulated) from the underlying theory, must be at least a 
nonlocal function(al) parametrized by the fundamental parameters of 
the underlying theory and must be closely related with the 
gravitation interactions and perhaps new fundamental ones, rather 
than a highly abstract Dirac delta function. In a sense, the 
incompleteness of the present QFTs or their ill-definedness is 
inherent in the present quantization procedure whose most elementary 
technical building block is Dirac function (called as distributions 
by mathematicians) that is \it extremely singular and can not be
defined in the usual sense of function. \rm That the distribution 
theory works necessarily with test function space or appropriate 
measure, if viewed from physical angle, is equivalent to that we need
more 'fundamental structures' in order for some singular functions
to make sense, i. e., a necessity of introducing underlying theory
or its artificial substitute--regularization. The constructive 
field theory approach, in this sense, also works with a 
regularization effected through the differential properties($C^{k}$) 
of the test functions. The author is not clear about the further 
implications of this observation yet.

In summary, we discussed the some important issues around a recently
proposed approach for renormalization which is simple to work with
and applicable to all QFT models in any space-time dimension. Some
related observations are given.

\vspace{3.0cm}
The author is grateful to Professor J. Polchinski for helpful discussions
and to Professor G.-j. Ni for continuing encouragement.

\vspace{1.0cm}


\begin{thebibliography}{99}
\bibitem {YYY} Jifeng Yang, Report No. hep-th/9708104.
\bibitem {Jac} R. Jackiw, Report No. hep-th/9709212.
\bibitem {DR}  see for example, G. 't Hooft and M. J. G. Veltman, Nucl. 
Phys. {\bf B 44}, 189 (1972); L. Culumovic \sl et al  \rm, Phys. 
Rev. {\bf D 41}, 514 (1990); D. Evens \sl et al \rm, Phys. Rev. 
{\bf D 43}, 499 (1991); D. Z. Freedman \sl et al \rm, Nucl. Phys. 
{\bf B 371}, 353 (1992); P. R. Mir-Kasimov, Phys. Lett. 
{\bf B 378}, 181 (1996);  J. J. Lodder, Physica {\bf A 120}, 1, 30 
and 508 (1983); H. Epstein and V. Glaser, Ann. Inst. Henri Poincare 
{\bf XIX}, 211 (1973). The Epstein-Glaser's constructive approach
also invokes the procedure equivalent to introducing regularization,
i.e., test function space or the interaction strength function, see
also later application of the E-G approach, A. Aste, G. Scharf and 
M. Dutch, J. Phys. A {\bf 30}, 5785 (1997) and references therein.
\bibitem {wein} S. Weinberg, \it The Quantum Theory of Fields, \rm
Vol. I, Ch. XII, Section 3, Cambridge University Press,1995. The author
is grateful to Professor J. Polchinski for this information.
\bibitem {wil} K. G. Wilson, Phys. Rev D {\bf 4}, 3174, 3184 (1971);
K. G. Wilson and J. G. Kogut, Phys. Rep. {\bf 12} 75(1974)
\bibitem {pol} J. Polchinski, Nucl. Phys. B {\bf 231}, 269(1984).
\bibitem {CK} W. E. Caswell and A. D. Kennedy, Phys. Rev. {\bf D 25}, 
392 (1982).
\bibitem {YP} Jifeng Yang, Ph D dissertation, Fudan University, 1994,
unpublished.
\bibitem {LL} C. H. Llewellyn Smith, Phys. Lett. {\bf B 46}, 233 (1973).
\bibitem {Y1}  Jifeng Yang, in preparation.
\bibitem {JH} J. H. Schwarz, Nucl. Phys. {\bf B} (Proc. Suppl.) 
{\bf 55 B},1 (1997); hep-th/9607201.
\bibitem {gom} J. Gomis and S. Weinberg, Nucl. Phys. B {\bf 469} 473 
(1996) and references therein.
\bibitem {JY} Jifeng Yang, in preparation.
\bibitem {Kino} T. Kinoshita,  J. Math. Phys. {\bf 3}, 650 (1962); 
T. Kinoshita and A. Ukawa, Phys. Rev. {\bf D 13}, 1573 (1976);
 T. D. Lee and M. Nauenberg, Phys. Rev. {\bf B 133}, 1549 (1964).
\bibitem {Haag} R. Haag, Local Quantum Physics, Springer-Verlag, 
(1993); M. Lavelle and D. McMullan, UAB-FT-369 and PLY-MS-95-03.
\bibitem {Acta} J-f Yang and G-j Ni, Acta Physica Sinica {\bf 4},
88 (1995); J-f Yang and G-j Ni, Report No.hep-th/9801004,
 to appear in Mod. Phys. Lett. A; 
G-j Ni and J-f Yang, Phys. Lett. {\bf B 393}, 79 (1997).
\end{thebibliography}
\end{document}